\documentclass[fleqn,12pt,twoside]{article}
\usepackage{espcrc1}

\usepackage{graphicx}
\newcommand{\AmS}{{\protect\the\textfont2
  A\kern-.1667em\lower.5ex\hbox{M}\kern-.125emS}}

\title{Baryons on the lattice}
\author{G.S. Bali\address{Dept.\ of Physics \& Astronomy, The University
of Glasgow, Glasgow G12 8QQ, UK}}
\begin{document}
% typeset front matter
\maketitle
\begin{abstract}
I comment on progress of lattice QCD techniques and calculations.
Recent results on pentaquark masses as well as of
the spectrum of excited baryons are summarizned and interpreted.
The present state of calculations of quantities related to
the nucleon structure and of electromagnetic transition form factors
is surveyed.
\end{abstract}
\section{INTRODUCTION}
In the past two years several narrow
hadronic resonances have been discovered: new bottomonium and
charmonium states, the $B_c$ and at least two new $D_s^*$ mesons.
It is more than twenty years ago that new states with widths $<10$~MeV
have been seen last, in the $\Upsilon$ system.
This is an exciting situation as only a rather small
number of discovered hadrons, including most notably quarkonia, the
hydrogen of QCD, are narrow. What makes this new era of hadron spectroscopy
even more interesting is that many of these previously overlooked
states appear to require more QCD than is allowed within
simple quark-model $q\bar{q}$ mesons or $qqq$ baryons:
at least
some of the new states constitute the anti-thesis to the ``hydrogen of QCD''.
If confirmed in a high statistics experiment, the $\Theta^+$ pentaquark baryon
and possibly other exotic baryons will add even more to this excitement.

Lattice QCD is ideally positioned to
compute the spectrum of reasonably stable hadrons as well as of
non-perturbative properties relating to their internal structure.
The above described experimental discoveries
were paralleled by significant advances in lattice methods,
enabling computer simulations of QCD to become a precision
predictive tool. In many areas there are also lessons to be learned
from combining lattice studies with effective field theory (EFT) methods
and/or model assumptions.

We have also witnessed experimental progress in the study of the
spectrum of excited baryons and of electromagnetic transition
form factors. Generalised parton distributions (GPDs)
of the proton
are now being studied intensively. As a stable particle the
nucleon lends itself to lattice studies. In the case of spin-independent
structure functions it will be hard for lattice simulations to
compete with the experimental precision.
However, in addition to the theoretical satisfaction of
verifying experimental measurements, this provides an ideal test ground
for the methods and approximations employed in lattice studies.
Once spin and transversity are included into the description
of the nucleon, the experimental situation is far less clean and
here there is real
potential for lattice prediction rather than postdiction.

I will describe the present state of the field, interpret recent
pentaquark and baryon mass calculations and briefly survey
studies on the form and
structure of baryons.
\section{THE LATTICE: WHERE ARE WE? WHERE DO WE GO?}
QCD can be regularized by introducing a space-time lattice cut-off $a$.
The QCD coupling and $n_f$ quark masses whose values are not predicted by
QCD should then be matched to reproduce $n_f+1$
experimental measurements of hadronic properties, for instance
hadron masses.
Everything else is a prediction and in this sense lattice QCD is
a {\em first principles} approach. The confinement of colour
implies that finite size effects are usually tiny, as long as
the spatial box extent $La\gg m_{\pi}^{-1}$.
We are fortunate to find that lattice spacings
$a^{-1}=1$ -- 4~GeV are sufficiently small
to allow for controlled continuum limit extrapolations,
$a\rightarrow 0$. This means that
$L\ll 100$ is sufficient, which makes QCD tractable on computers.

\begin{figure}[tb]
\begin{minipage}[t]{85mm}
  \includegraphics*[height=.245\textheight]{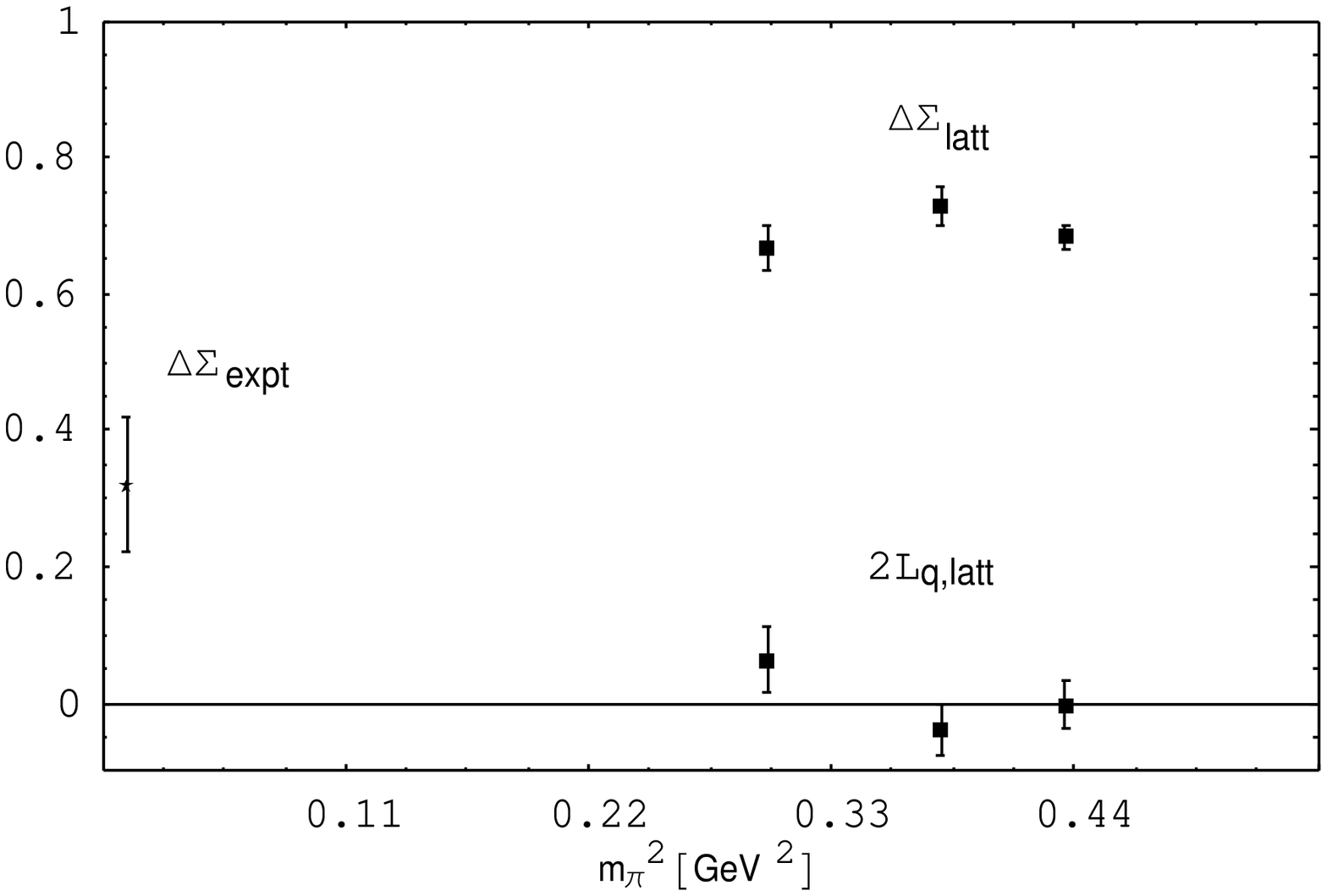}\\[-1.3cm]
  \caption{The fraction of the nucleon spin carried by its quarks~\cite{Negele:2004iu}.}
\label{fig:spin}
\end{minipage}
\hspace{\fill}
\begin{minipage}[t]{70mm}
\includegraphics[height=.26\textheight]{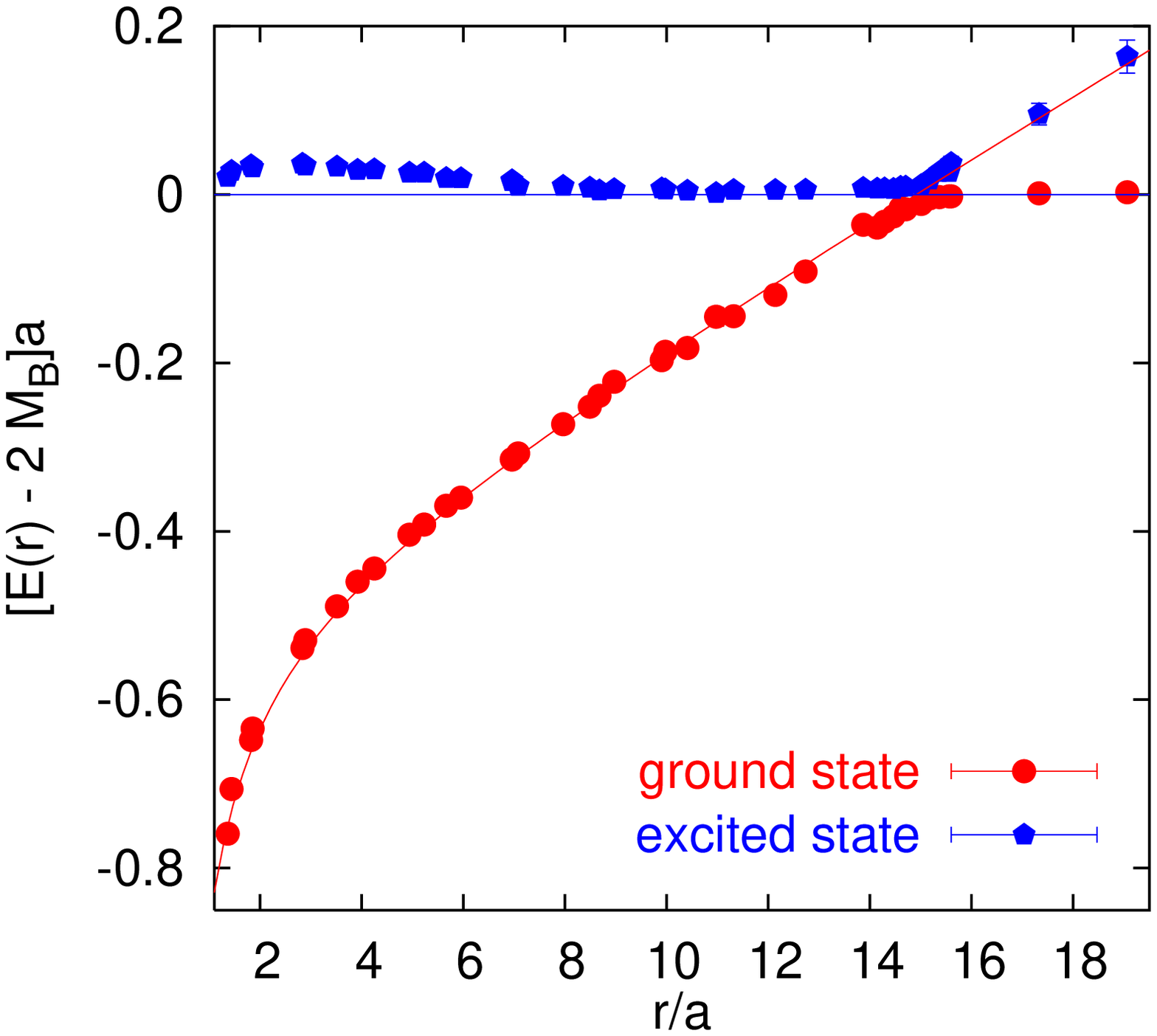}\\[-1.3cm]
\caption{The lowest two energy levels in the $Q{\overline Q}$ system
with sea quarks~\cite{Bali:2004pb}.}
\label{fig:break}
\end{minipage}
\end{figure}
On a lattice with $V=L^3T$ points, the lattice Dirac operator is a
huge matrix of dimension $12V$. The inversion of this operator
represents the major computational task of lattice QCD and this makes
simulations incorporating sea quarks expensive.
The algorithmic cost explodes
with small $\pi$ masses, $\propto1/(m_{\pi}a)^{\simeq 3}$. A smaller
$m_{\pi}$ also requires a larger spatial lattice volume and the 
scaling behaviour, keeping
$m_{\pi}La$ fixed, is even worse: $\propto 1/(m_{\pi}a)^{\simeq 7}$.
This is the main reason why many simulations are performed in the
quenched approximation including only valence quarks and
neglecting the polarization of the QCD vacuum
due to sea quarks. While this approximation violates unitarity and
does not even qualify as a quantum field theory, light hadron masses
seem to agree within 10~\% with
experiment~\cite{Gattringer:2003qx}, indicating that the
main effect of quark loops can be absorbed into redefinitions of the bare
parameters of the theory. The quenched approximation however goes terribly
wrong at least in the scalar and pseudoscalar flavour-singlet sectors.

In Figure~\ref{fig:spin}
we display recent $n_f=2$ QCD results obtained by the LHP and SESAM
Collaborations~\cite{Negele:2004iu}
on the quark contribution $\Delta\Sigma$ to the proton spin,
in the $\overline{MS}$ scheme at a scale $\mu=2$~GeV.
The normalization is such that $\frac12=\frac12\Delta\Sigma+L_q+J_g$,
where $L_q$ is the contribution from the quark angular momentum and
$J_g$ from the gluons. Similar results have been obtained by
the QCDSF Collaboration~\cite{Gockeler:2003jf}.
In these simulations, $m_{\pi}>550$~MeV.
Obviously, for infinite quark masses we expect $\Delta\Sigma=1$. It
is therefore not surprising that the experimental value is overestimated
and it is clear that smaller quark masses
are absolutely essential to allow for a meaningful chiral extrapolation.
Fortunately, with the advent of new Fermion
formulations~\cite{Neuberger:1997fp} that
respect an exact lattice chiral symmetry,
a reduction of the quark mass towards
values $m_{\pi}\approx 180$~MeV has become possible~\cite{Dong:2003zf},
albeit so far only in the quenched approximation.

In many lattice calculations it is sufficient to calculate quark propagators
that originate from a fixed source point. 
In these cases only one
column of the inverse Dirac matrix needs to be calculated,
na\"{\i}vely reducing the
effort by a factor $V.$
In some cases diagrams with disconnected quark lines are needed.
Examples are the physics
of flavour singlet mesons, strong decays as well as
parton distributions. In the latter case the complication can
be avoided by
assuming $SU(2)$ isospin symmetry and
only calculating differences between $u$ and $d$ quark distributions.
Disconnected quark lines require all-to-all propagators and hence 
a complete inversion of the Dirac matrix appears necessary.
This turns out to be prohibitively expensive in terms of memory
and computer time. Fortunately,
sophisticated noise reduced stochastic estimator
techniques have been developed over the past few years and as a result
tremendous progress was achieved. One such benchmark is the
QCD string breaking
problem, $Q(r)\overline{Q}(0)\leftrightarrow\overline{B}(r)B(0)$, where
$B=\overline{Q}q$ and $Q$ is a static quark. This represents
one of the cleanest examples of a strong decay.
Within both, the transition matrix element as well as the $\overline{B}B$
state all-to-all propagators are required.
In Figure~\ref{fig:break} we display the result of a recent SESAM Collaboration
study~\cite{Bali:2004pb} with $n_f=2$, $m_q\approx
m_s$ and $a\approx 0.085$~fm.
An extrapolation to physical light quark masses yields a string breaking
distance $r_c\approx 1.16$~fm.
The gap between
the two states in the string breaking region is
$\Delta E\approx 50$~MeV and we are
able to resolve this with a resolution of 10 standard deviations!

In conclusion, all the long standing ``killers'' $m_q\ll m_s$, $n_f>0$ and
all-to-all propagators have been successfully tackled. However, we are still
a few years away from overcoming combinations of two of these simultaneously
and possibly up to a decade separates us from precision simulations of
flavour singlet
diagrams with realistically light sea quarks.
Not all these ingredients are always required at the same time.
While ever bigger computers are an absolute necessity, most of the
recent progress
would have been impossible without novel methods.
The gain factor from
faster computers was almost 5,000 over the past 15 years.
The factor from theoretical and algorithmic advances is harder to quantify.
\section{PENTAQUARKS}
QCD goes beyond the quark model and hence
hadronic states that do not fit into a na\"{\i}ve quark model of
$q\bar{q}$ mesons and $qqq$ baryons are of particular interest.
Many of the observed hadrons will contain considerable higher
Fock state components. Obviously,
quantum numbers that are incomprehensible with a quark model meson
or baryon interpretation provide us with the most 
clean-cut distinction. Such examples do exist in the Review of Particle
Properties, namely the $J^{PC}=1^{-+}$ mesons $\pi_1(1400)$
and $\pi_1(1600)$. The minimal configuration required to obtain a vector
state with positive charge either consists of two quarks and two antiquarks
(tetraquark/molecule) or of quark, antiquark and a gluonic excitation
(hybrid meson). These resonances are rather broad
with widths $\Gamma\approx$~300~MeV.
However, the ratio $\Gamma/m$ is very much the same as for the
established $\rho(770)$ vector meson.

Another clear indication of an $n_{\rm quark}>3$
nature would for instance
be a baryonic state with strangeness $S=+1$. The minimal quark configuration
in this case consists of five quarks (pentaquark): $uudd\bar{s}$.
Over the past two years several experiments
have presented evidence of a very narrow
$\Theta^+(1530)$ resonance~\cite{polya}, with decay
$\Theta^+\rightarrow pK^0$ and $\Theta^+\rightarrow nK^+$.
The parity has not yet been established. However the mass is
about 100~MeV above the $KN$ threshold and for $J^P=\frac{1}{2}^-$
an $S$-wave decay is possible, which might be difficult to reconcile with
a width $\Gamma \ll 10$~MeV. For $\frac{1}{2}^+$ a $P$-wave is required, still
a bit puzzling but less so. As the main decay channel does not require
quark-antiquark pair creation one might hope to gain some
insight from quenched lattice simulations
and several attempts have been
made~\cite{Csikor:2003ng,Sasaki:2003gi,Chiu:2004gg,Mathur:2004jr,Ishii:2004qe,Alexandrou:2004ws}.

Two groups~\cite{Sasaki:2003gi,Chiu:2004gg} also investigated charmed
pentaquarks and two
studies~\cite{Csikor:2003ng,Mathur:2004jr} incorporated the $I=1$ sector,
in addition to $I=0$. Two groups~\cite{Chiu:2004gg,Mathur:2004jr}
employed chiral overlap Fermions while the others used conventional
Wilson-type lattice quarks. Only the 
Kentucky group~\cite{Mathur:2004jr} went down to $m_{\pi}\approx 180$~MeV
while all other $\pi$ masses were larger than 400~MeV. The
Budapest-Wuppertal group~\cite{Csikor:2003ng} varied the lattice spacing and
attempted a continuum-limit extrapolation.
In all studies the negative parity mass came out to be lighter than the
positive parity one, which is expected in the heavy quark limit.

There are two crucial questions to be asked: what happens when
realistically light quark masses are approached? Do we see
resonant or scattering states?
Resolving a resonance sitting on top of a tower of
$KN$ scattering states with different relative momenta appears rather
hopeless at first. However, there are two discovery tools available:
variation of the lattice volume and of the creation
operator.
By varying the volume (and the boundary
conditions~\cite{Chiu:2004gg})
one will change the spectrum of $KN$ scattering states
as well as the coupling of a given operator to $KN$ (the spectral weight).

For the $\frac{1}{2}^+$ state which can
only decay into a $P$-wave, the mass of the scattering
state will depend on the lattice size since the smallest possible non-vanishing
lattice momentum is $\pi/(aL)$. For $\frac{1}{2}^-$ the volume dependence of the
lowest scattering state mass will be weak, however, the scaling of the
spectral weight with the volume provides us with additional
information.

\begin{figure}
\includegraphics[height=.25\textheight]{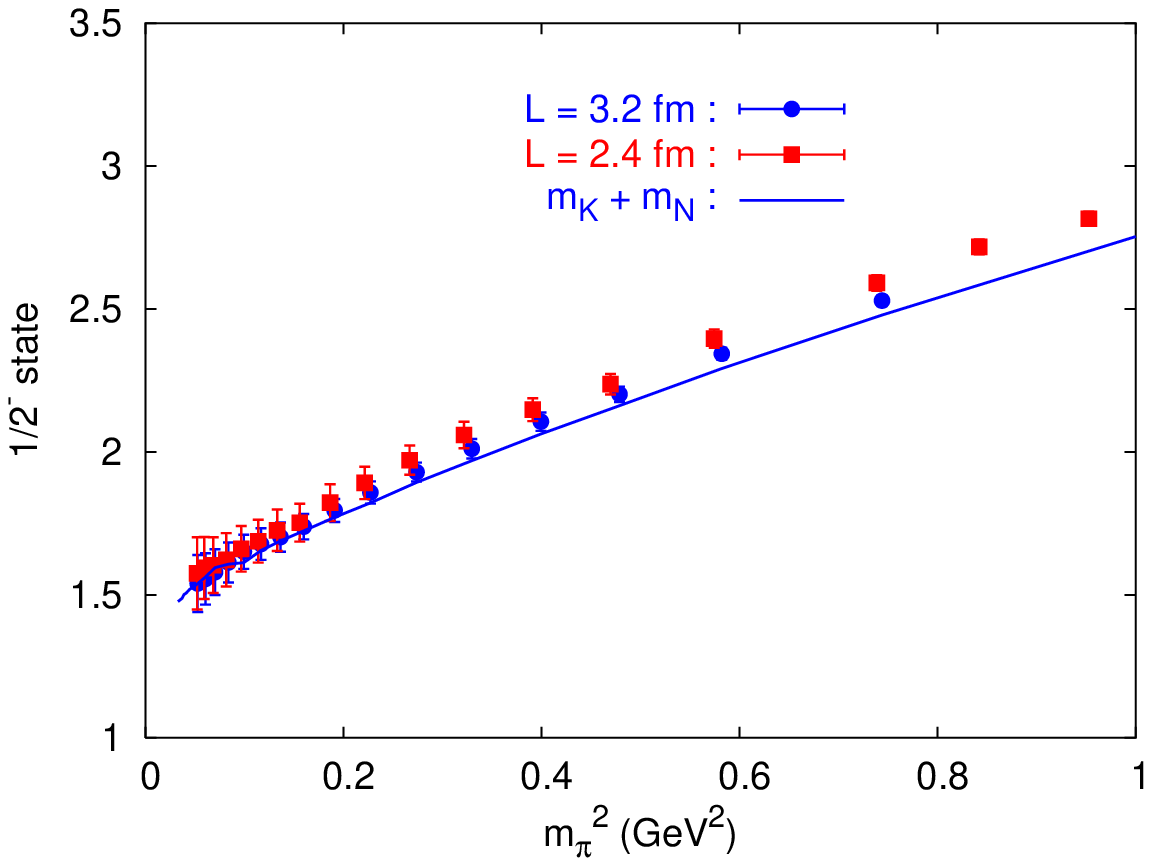}\hspace{\fill}
\includegraphics[height=.25\textheight]{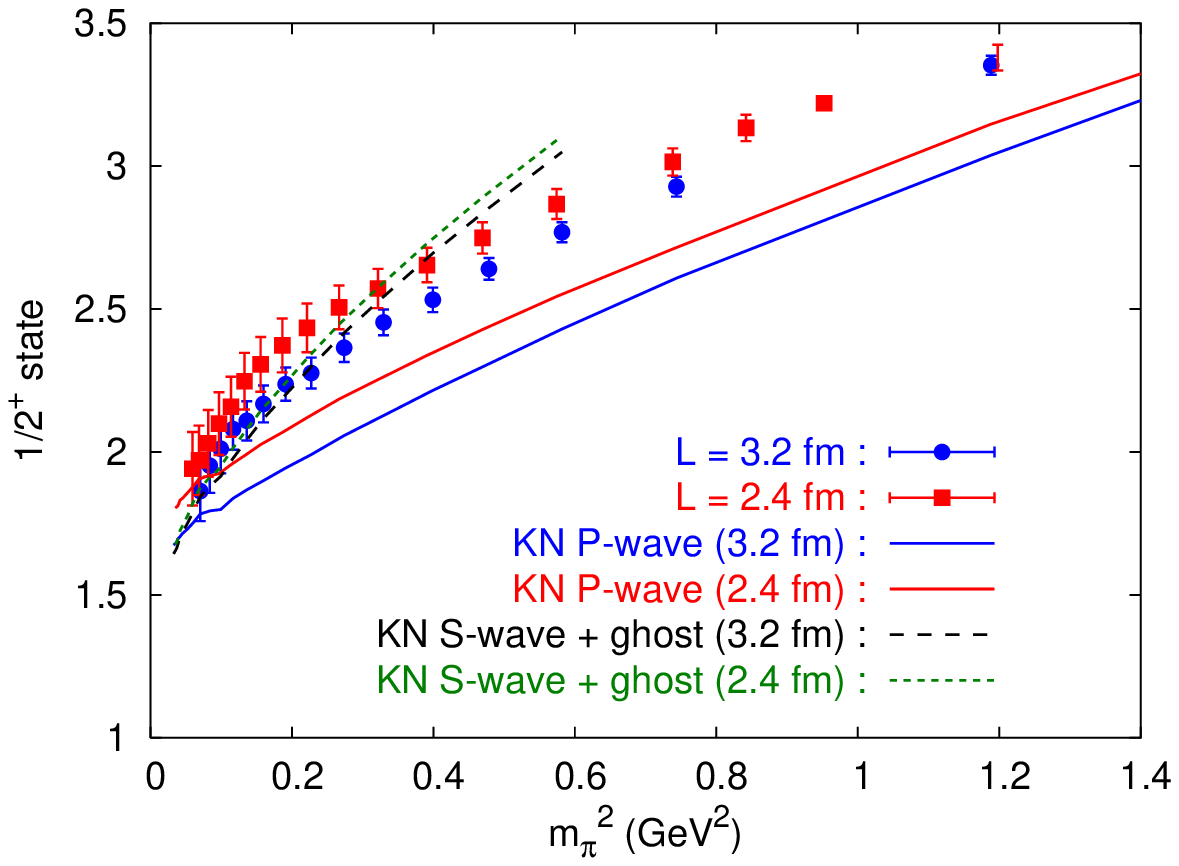}\\[-1.3cm]
\caption{The $\pi$ mass dependence of $I=0$
$uudd\bar{s}$ ``pentaquark'' masses~\cite{Mathur:2004jr}.}
\label{fig:penta}
\end{figure}
It turns out that the situation on the lattice is at least as ambiguous as
the one encountered in experiment. To demonstrate this we display some
Kentucky-Washington results~\cite{Mathur:2004jr} in Figure~\ref{fig:penta}.
It appears that the $\frac{1}{2}^-$ state dominantly couples to an
$S$-wave $KN$. The $\frac{1}{2}^+$ displays the qualitative volume dependence
of a $P$-wave, however, it does not share its non-interacting mass.
It is most likely a $P$-wave scattering state. This interpretation is
supported by the observed volume dependence of the
spectral weight. At very small $m_{\pi}$ the situation
becomes further complicated by the fact that there is no axial anomaly in the
quenched approximation. Hence the flavour singlet $\eta'$ is degenerate
with the $\pi$. As this contribution comes in with a negative spectral weight,
it is sometimes labelled a ``ghost''. The $\frac{1}{2}^+$ state can contain
such a $KN\eta'$ $S$-wave (dashed lines).

The $\frac{1}{2}^-$ state
becomes indistinguishable from a $KN$ $S$-wave as the quark mass is reduced.
This does not conclusively exclude the possible existence of a nearby resonant 
state which might only couple very weakly to the creation operator used
in this particular study. In order to draw more definite conclusions
a variation of the creation operator as well as of the volume appears
necessary, which is a very ambitious project~\cite{Fleming:2005jz}.
If the pentaquark really was such a narrow resonance
as some experiments suggest then maybe a lattice operator can be constructed
that has a large overlap with this state but only a very small coupling
to $KN$.

Lattice studies of diquark interactions 
in a simplified, more controlled environment represent
an alternative strategy to the brute force simulation of unstable states.
A baryon with one static and two light quarks constitutes one such arena.
One can of course also investigate multiquark interactions
in the nonrelativistic
limit of infinitely heavy quark masses. Such tetra- and penta-quark potentials
have been studied recently by two
groups~\cite{Alexandrou:2004ws,Alexandrou:2004ak,Okiharu:2004wy}
and the results
should provide model builders with some insight. However, it is not
clear how to relate these findings to the light quark limit in which
chiral symmetry appears to play a bigger r\^ole than instantaneous
confining forces.

There exist quite a few narrow resonances very close to strong decay thresholds
like the $\Lambda(1405)$, the recently discovered $X(3872)$ charmonium state
and the $a_0/f_0(980)$ system. It is very conceivable that such states
contain a sizable multiquark component. The question then arises
if these are would-be quark model states or if these are
true molecules/multiquark-states, that appear {\em in addition}
to the quark model
spectrum. A fantastic arena to
address this was provided by the recently discovered (probably scalar)
$D_s^*(2317)$ and (probably axialvector) $D_s^*(2457)$ states.
First lattice studies~\cite{Bali:2003jv,Dougall:2003hv}
have been performed,
with somewhat contradictory interpretations of very compatible results.
One might hope that a similar lattice effort will also be dedicated on
the comparatively cleaner and easier question of tetraquarks as
has been on pentaquarks.
\section{EXCITED BARYONS}
The spectrum of baryons has attracted renewed experimental and theoretical
interest in recent years. There is the question if the states can be
understood in terms of quark models and if so by what sort of
interaction and assumptions.
Do gluonic excitations or pentaquark components play a r\^ole
for instance in the Roper resonance? What can we learn about
quark-quark interactions within bound states? Quark model predictions
are somewhat obscured as the corresponding decay widths set a limit
on the precision that can be expected for the resulting masses.
Are missing states really ``missing'' or are they just obliterated
due to the presence of many very broad, overlapping resonances?
Strongly decaying hadrons also pose problems in lattice simulations.
At present almost all calculations of baryonic resonances
have been performed within the quenched approximation in which these
are stable and hence the problem is circumvented.

This limitation can also be viewed as a virtue since most models suffer from an
omission of quark pair creation
effects too. Comparison with similar lattice results
then allows to establish the validity range and applicability of
a particular phenomenological model. One strength of lattice
methods is that simulations are not limited to the quark mass
parameters found in nature. Investigating the quark mass dependence of results
is a powerful tool. In the limit of large quark masses one would
expect to make contact with non-relativistic quark models while
as $m_{\pi}\rightarrow 0$ overlap with chiral perturbation theory ($\chi$PT)
predictions should be verifiable.

Based on the assumptions that QCD bound states are mesons and
baryons, that there is a mass gap and spontaneous chiral
symmetry breaking at zero quark mass, an effective low energy chiral
effective field theory ($\chi$EFT) can be derived in the spirit of the
Born-Oppenheimer approximation.
This will, to leading order,
describe interactions between the (fast moving)
massless Nambu-Goldstone pions and
other hadrons. In nature quarks and thus pions are not massless
and the leading mass corrections are formally
of order $m_{\pi}/\Lambda_{\chi SB}$ where $\Lambda_{\chi SB}\approx
4\pi F_{\pi}>1$~GeV. The
number of terms explodes at higher orders and predictive power is eventually
lost, unless $m_{\pi}$ is sufficiently small to allow for
an early truncation.

Lattice simulations with sea quarks have so far been limited
to unrealistically heavy
pions, heavier than about 400~MeV. Only
recently masses as low as 180~MeV
have become possible~\cite{Dong:2003zf}.
To allow for a controlled extrapolation of lattice results to
the physical region it is mandatory
to establish an overlap between simulation data
and $\chi$PT expectations.
In general the size of this window will depend on the observable in question.
Chiral lattice Fermion actions will make such a comparison with
$\chi$PT cleaner. Only in this case an exact
version of chiral symmetry can be formulated at finite lattice spacing.
With other Fermion discretizations, strictly speaking,
a comparison should only be attempted
after extrapolating lattice results to the continuum limit.
While even $400$~MeV~$\ll\Lambda_{\chi SB}$ (modulo the ambiguity of ``$\ll$''
 {\em vs.}\/
``$<$'') such a pion is still doomed to ``see'' some of the internal
structure of the proton, with an inverse charge radius of about 250~MeV.
Hence it is doubtful that the quark and gluon
nature of QCD can completely be ignored with such a ``hard'' pion probe.

Na\"{\i}vely, hadron masses are a polynomial in
the quark mass, $m_q\propto m_{\pi}^2$. However, pion
loops give rise to a non-analytic functional dependence on the quark
mass. For instance the nucleon
mass is given by,
\begin{equation}
\label{eq:fit}
m_N(m_{\pi})=m_N(0)+a_2m_{\pi}^2+a_3m_{\pi}^3+\left[e_1^r(\lambda)+a_4
+a_4'\ln\frac{m_{\pi}}{\lambda}\right]m_{\pi}^4+a_5m_{\pi}^5+\cdots,
\end{equation}
with a renormalization scale $\lambda$. 
The coefficients of the non-analytic terms can be related
to phenomenological 
low energy constants. For instance $a_3=-3g_A^2/(32\pi F_{\pi}^2)$.
Apart from such constants, $a_4$ and $a_4'$
contain terms $\propto m_N^{-1}$ and $a_4'$ a contribution $\propto a_2$.
In the quenched approximation,
the leading non-analytic term is not proportional
to $m_{\pi}^3$ but to $m_{\pi}^2\ln(m_{\pi}/\lambda)$, due to the
$\eta'$ becoming an additional Goldstone pion.

\begin{figure}[th]
\centerline{\includegraphics[height=.3\textheight]{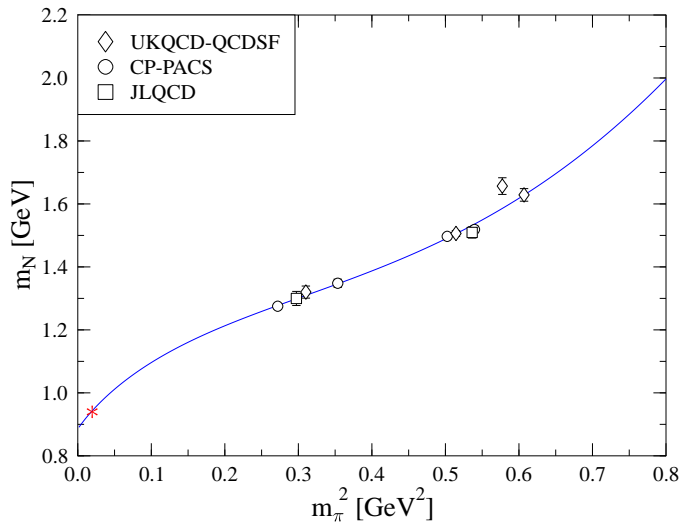}}
\vspace*{-1cm}
\caption{Chiral extrapolation of the nucleon mass~\cite{procura,meinulf}.}
\label{fig:meinulf}
\end{figure}
In Figure~\ref{fig:meinulf} we show a comparison~\cite{procura,meinulf}
between $n_f=2$ lattice data of the nucleon
mass and the  relativistic
$\chi$PT expectation Eq.~(\ref{eq:fit})~\cite{procura}.
The lattice results were obtained by the CP-PACS and JLQCD
Collaborations~\cite{AliKhan:2001tx} as well as by the UKQCD and QCDSF
Collaborations~\cite{Allton:2001sk},
with (non-chiral) Wilson-type Fermions,
on relatively fine lattices, $a<0.15$~fm and large
volumes, $aLm_{\pi}>5$.
The low energy parameters 
were fixed to phenomenological values and the
fit comprises only of $m_N(0),c_2$ and $e_1^r(1\,\mbox{GeV})$.
The quantitative agreement between curve and data
for $m_{\pi}>600\,\mbox{MeV}$ is accidental~\cite{procura}.
A na\"{\i}ve polynomial fit to the simulation data
results in a nucleon mass
much larger than the experimental value:
lower order $\chi$PT only becomes applicable at smaller quark masses.

EFTs are based on the separation of scales.
If the $\chi$EFT however is regulated in dimensional regularization
then $\pi$ loop-integrals can receive significant contributions
from momenta $q> \Lambda_{\chi SB}$. To enhance the convergence
the Adelaide group~\cite{Young:2002cj}
suggested a ``finite range regularization'' approach which
amounts to introducing a momentum cut-off which is then varied
to achieve ``model independence''. A hard
cut-off can also be provided by lattice regularization of the chiral
expansion~\cite{Shushpanov:1998ms,Lewis:2000cc}. Needless to say
that all cut-off and scheme dependence will disappear at sufficiently
high orders in the $p$-expansion.

During the past four years we witnessed many
lattice publications
on the spectrum of excited baryons~\cite{Leinweber:2004it}.
While the extraction of a ground state mass is relatively straight
forward, radial excitations either require
high statistics and some confidence in the fitting procedures or
the design of sophisticated, non-local creation operators~\cite{Burch:2004he}.
All but one study~\cite{Maynard} have been performed in the
quenched approximation.
Only the LHPC-UKQCD-QCDSF Collaborations~\cite{Gockeler:2001db},
using the Wilson-clover action,
attempted a continuum limit extrapolation.
Other strategies were implementations of improved Wilson-type
actions like the $D_{234}$ action by Lee {\it et al.}~\cite{Lee:2001ts}
or the FLIC action by Melnitchouk {\it et al.}~\cite{Melnitchouk:2002eg}
as well as a recent study with Wilson-clover Fermions~\cite{Guadagnoli:2004wm}.
The BGR Collaboration employed chirally improved
Fermions~\cite{Burch:2004he}, the Riken-BNL group used chiral
domain wall Fermions~\cite{Sasaki:2001nf} and the Taiwan~\cite{Chiu:2005zc}
as well as the Kentucky-Washington groups~\cite{Dong:2003zf} made use
of overlap Fermions.

Early articles shared the observation of
the positive parity state being much heavier than the
Roper $N'(1440)$ resonance while the negative parity ground state
was compatible with the orbital excitation, $N^*(1535)$.
One explanation would be that
the resonance observed in nature might have little overlap with the
dominantly $qqq$ state created on the lattice; pentaquark
or gluon components might be necessary. Alternatively,
maybe one should not take the exact position of a resonance
with a width of $O(200\,\mbox{MeV})$ overly seriously.

\begin{figure}[htb]
\begin{minipage}[t]{78mm}
\includegraphics*[width=.99\textwidth]{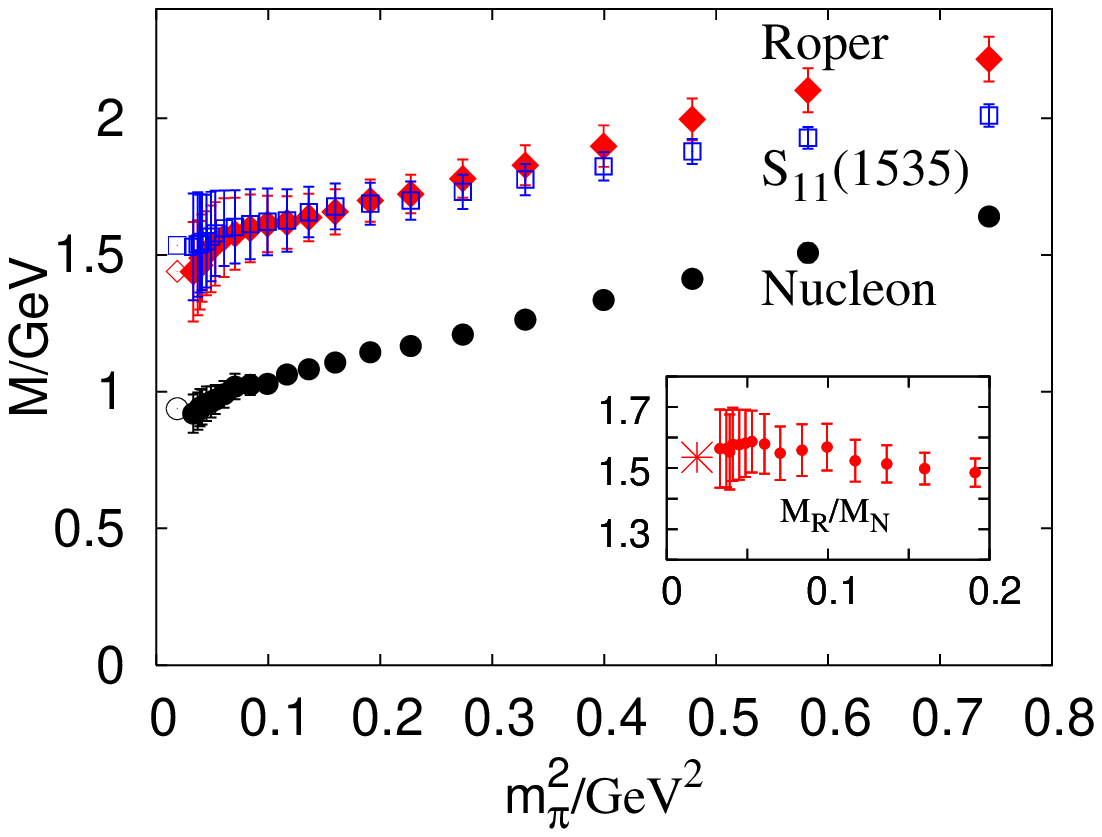}\\[-1.5cm]
\caption{The $N$, $N'$ and $N^*$ masses for overlap
Fermions~\cite{Dong:2003zf}
at $a\approx 1$~GeV~.}
\label{fig:dong}
\end{minipage}
\hspace{\fill}
\begin{minipage}[t]{78mm}
  \includegraphics*[width=.92\textwidth]{M_vs_m.eps}\\[-1.5cm]
  \caption{Chirally improved Fermions~\cite{Burch:2004he}:
$0.06<m_{\pi}^2/\mbox{GeV}^2<0.65$, $a^{-1}\approx 1.33$ GeV.}
\label{fig:lang}
\end{minipage}
\end{figure}
New results obtained by the Kentucky-Washingon
group~\cite{Dong:2003zf} and the BGR Collaboration~\cite{Burch:2004he}
at lighter pion masses, $m_{\pi}>180$~MeV and $m_{\pi}>250$~MeV,
respectively, are however compatible with experiment. We display these
in Figures~\ref{fig:dong} and \ref{fig:lang}. Rather unsurprisingly
the creation operator requires a node in its spatial wave function
to produce a significant overlap with the radial excitation.
This complication was circumvented in
Ref.~\cite{Dong:2003zf} by a sophisticated fitting procedure while
in Ref.~\cite{Burch:2004he} such an adequate operator has been
constructed, using
a variational principle. Making contact with the light quark regime
seems hopeless for
$m_{\pi}>500$~MeV. However, the authors of
two recent studies~\cite{Guadagnoli:2004wm,Chiu:2005zc}
manage to extrapolate their results to the experimental
values from such pion masses as well.
Note that the BGR Collaboration only sees a
clear signal of the radial excitation for
$m_{\pi}>400$~MeV while the Bayesian fitting procedure of
the Kentucky group yields results at any quark mass.
The Taiwan group~\cite{Chiu:2005zc} in addition predicts the spectrum
of doubly charmed baryons, with findings roughly compatible
with earlier studies~\cite{Lewis:2001iz,Flynn:2003vz} as well as with the
SELEX candidate(s). 

One might hope that in the near
future the $N'$ state can cleanly be disentangled from a $P$-wave $\eta'N$
scattering state or, at very small masses, from $S$-wave
$\pi\pi N$/$\eta'\eta' N$ states.
By studying the volume dependence and spectral weights
the Kentucky group
has taken steps in this direction.
Most of the lattice studies
include additional baryonic resonances where similar problems need
to be addressed.
In conclusion, we are close to an understanding of the transition between
the heavy and light quark limits in the quenched approximation,
an information invaluable to model builders.

\section{FORM AND STRUCTURE}
Quite a few results on moments of GPDs, most notably
from the QCDSF~\cite{Gockeler:2004vx}, SESAM and
LHP Collaborations~\cite{Hagler:2003jd}, exist.
A nice review of the state-of-the-art concerning spin-independent
parton distributions and the axial charge can be found in
Ref.~\cite{Schroers:2005rm}.
The main problem here is an overestimation by about 60~\%
of $\langle x\rangle_{u-d}$ if linearly extrapolated
in $m_{\pi}^2$, relative to experiment. It
is not clear whether this difference will reduce as the quark mass
is decreased or if this has to do with the
non-chiral Fermion formulation used. This issue will be clarified in
the near future~\cite{Gurtler:2004ac}.

There has also been progress in resolving the momentum dependence
of electromagnetic $\gamma^* N\rightarrow\Delta$ transition form
factors in a quenched study~\cite{Alexandrou:2004xn},
in the region $0.1\,\mbox{GeV}^2<Q^2<
1.4\,\mbox{GeV}^2$. The magnetic dipole form factor is significantly
overestimated at large $Q^2$, due to the unrealistically small
charge radii of $\Delta$
and $N$. One might hope that such effects cancel in part from
form factor ratios. $R_{EM}=G_{E2}/G_{M1}$
is fairly constant at -0.02(1), once extrapolated to the chiral limit.
In contrast, $R_{CM}=G_{C2}/G_{M1}$ decreases monotonously from 
-0.01(1) at 0.1~GeV${}^2$ down to -0.09(3) at $Q^2>1\,\,\mbox{GeV}^2$.
This behaviour is
in good agreement with $Q^2>0.4\,\,\mbox{GeV}^2$
CLAS data while it
is hard to reconcile with the OOPS point
$R_{SM}=[-6.1\pm 0.2\pm 0.5]\,\%$ at $Q^2\approx 0.13\,\mbox{GeV}^2$.
\section{SUMMARY}
Many lattice studies now
include sea quarks. Within the quenched approximation, light quark masses close
to the physical limit have been realised and the lattice provides
a powerful tool for exploring the validity range of chiral expansions.
Lattice pentaquark studies still yield ambiguous results. A systematic
study using a large set of creation operators, of lattice volumes,
spacings and quark masses is possible but ambitious.
Latest lattice data
suggest that the mass of the Roper resonance can be reproduced in the quenched
approximation, thus indicating a non-exotic
leading Fock component. A lot of progress has been made in
understanding the structure of the nucleon and there is a strong
push towards reducing the lattice quark masses, closer to the physical limit.
\section*{ACKNOWLEDGMENTS}
I thank Thomas Hemmert for useful comments.
This work is part of the
EC Hadron Physics I3 Contract No.\ RII3-CT-2004-506078.
GB is supported by a PPARC Advanced
Fellowship (grant PPA/A/S/2000/00271) as well as by PPARC grant
PPA/G/0/2002/0463.

\end{document}